\documentclass[twocolumn,prb,showpacs,multicol,amsmath,amssymb]{revtex4-1}

\usepackage{graphicx}
\usepackage{dcolumn}
\usepackage{bm}
\usepackage{graphics}
\usepackage{soul}
\usepackage{epsfig,color}

\newcommand{\be}{\begin{equation}}
\newcommand{\ee}{\end{equation}}
\newcommand{\bea}{\begin{eqnarray}}
\newcommand{\eea}{\end{eqnarray}}

\usepackage[breaklinks,colorlinks,bookmarks=false,citecolor=red,linkcolor=blue,urlcolor=magenta]{hyperref}

\graphicspath{{figures/}} 

\begin{document}
\title[J. Vahedi]{Optimal performance of voltage-probe quantum heat engines}

\author{Zahra Sartipi} 
\affiliation{Department of Physics, Sari Branch, Islamic Azad University, Sari 48164-194, Iran}

\author{Javad Vahedi}
\email[]{javahedi@kip.uni-heidelberg.de}
\affiliation{Kirchhoff-Institut f\"ur Physik, Universit\"at Heidelberg, Im Neuenheimer Feld 227, 69120 Heidelberg, Germany}
\affiliation{Department of Physics, Sari Branch, Islamic Azad University, Sari 48164-194, Iran}

\date{\today}
\begin{abstract}
The thermoelectric performance at a given output power of a voltage-probe heat engine, exposed to an external magnetic field, is investigated in linear irreversible thermodynamics. For the model, asymmetric parameter, general figures of merit and efficiency at a given output power are analytically derived. Results show a trade-off between efficiency and output power, and we recognize optimum-efficiency values at a given output power are enhanced compared to a B\"uttiker-probe heat engine due to the presence of a characteristic parameter, namely $d_m$. Moreover,
similar to a B\"uttiker-probe heat engine, the universal bounds on the efficiency are obtained, and the efficiency at a given output power can exceed the Curzon-Ahlborn limit. These findings have practical implications for the optimization of realistic heat engines and refrigerators. By controlling the values of the asymmetric parameter, the figures of merit, and $d_m$, it may be possible to design more efficient and powerful thermoelectric devices.
\end{abstract}
\maketitle
\section{Introduction}\label{sec1}
The thermoelectric performance of quantum heat engines is an attractive field of interest for researchers from the experimental and theoretical communities. This is due to advances in material science and the constant demand for more advanced and powerful energy harvesting\cite{a1,a2,a3,a4,a441,a4433,a446,a551,vahedi2022}. Nevertheless, the efficiency of heat engines is restricted from above by the efficiency of Carnot heat engines. Therefore, fabricating  heat engines with high performance is highly demanded.
\par
It is well known that ideal heat engines working under reversible processes lack practical function due to their zero output power. Accordingly, finite-time thermodynamics was invented to optimize the thermoelectric performance of heat devices \cite{a443,a400,a411,a412,a413}. Moreover, macroscopic or quantum heat engines should not operate in the maximum output power region due to the relatively small efficiency, and thermal machines with higher efficiency can be achieved in a regime with  smaller output power. Whitney\cite{a35} showed that maximum efficiency for quantum heat engines can be obtained somewhere close to the maximum output power. Holubec and Ryabov \cite{a41,a42,a43} reported on the optimal performance of irreversible and low-dissipation heat engines. Based on their work, quantum heat engines performing near maximum output power can offer significantly higher efficiency than the maximum output power efficiency. Long et al.\cite{a44} investigated the efficiency of minimally nonlinear irreversible quantum heat engines at a given output power. This gave additional insight into efficiency since the output power of actual heat engines is below the maximum output power. 
\par
The optimization of real thermoelectric devices at a given power is of significant current interest due to the above-mentioned points. In this context exposing heat engines to an external magnetic field introduces complexity by breaking time-reversal symmetry in the system. Based on the work of Benenti et al.\cite{a5}, in any heat engine with broken time-reversal symmetry, both maximum efficiency and maximum output power efficiency are described by an asymmetry parameter and a general figure of merit. This could lead to an efficiency enhancement of the heat engine in the linear response regime. While the exact details of this process are still being studied and optimized\cite{a7,a8}, the idea of using external magnetic fields to enhance the efficiency of quantum heat engines is an active area of research with promising results\cite{a9}.
\par
In this context, adding more terminals to the heat engines also opens up the path to achieving higher efficiency and unveiling novel phenomena\cite{a11,a12}. Multiterminal heat engines offer several advantages over traditional two-terminal heat engines\cite{a310,a13,a18,a19,a2012,a2013}. For example, they can potentially achieve higher efficiency by allowing for more complex and efficient heat exchange between the engine and the external environment\cite{a14,a15,a16}. They can also exhibit novel thermodynamic phenomena that are not present in two-terminal heat engines. For instance, it might be possible for these complex systems to decouple charge and energy flows, leading to higher thermoelectric properties \cite{a17,a201,vahedi2018,vahedi2018_}. 
\par
For a multi-terminal device with an external magnetic field, due to the broken time-reversal symmetry, a new bound on the Onsager coefficients has been achieved within the framework of linear irreversible thermodynamics, offering a unique insight into the optimal performance of a heat engine \cite{a13,a18,a19,a310}. Zhang et al.\cite{a45} addressed how an external magnetic field could enhance the thermoelectric properties of a three-terminal quantum heat engine with a B\"uttiker probe (in which the probe reservoir is set to block both heat and particle currents) at a given output power. They demonstrated that broken time-reversal symmetry gives rise to a broad region of parameters for optimizing the performance of nano-scale heat engines. Furthermore, they obtained a different universal bound on the efficiency at a given output power of the quantum heat engine with broken time-reversal symmetry. In another work, Lu et al.\cite{a40} examined the thermoelectric properties of a  three-terminal thermoelectric engines with two independent output electric currents and one input heat current. Based on their work, the heat engine with two output electric currents can significantly increase efficiency and output power.

Despite the studies mentioned above, the performance optimization of a voltage-probe quantum heat engine \cite{a443, Gorini2021, Saha2022}, 
 ( which the probe reservoir is adjusted to only exchange energy but no charge current with the scattering region, as depicted in Fig.~\ref{f1}) with broken time-reversal symmetry has not been investigated yet. In the present work, we consider the voltage-probe setup and present detailed calculations and analyses of its performance at a given output power. In this regard, we need to mention that the most closely related previous work is that of Zhang et al.\cite{a45} who investigated a three-terminal setup in the presence of an external magnetic field in the context of B\"uttiker-probe.
\par
The rest of the paper is organized as follows. In Sec. \ref{sec2}, we elaborate on the efficiency properties at a given output power of voltage-probe quantum heat engines with an external magnetic field and show mathematically under which condition the voltage-probe heat engine is converted into the B\"uttiker-probe one. We present our main results in Sec. \ref{sec3}. Conclusions are summarized in Sec.\ref{sec4}. In appendixes.\ref{apeA}-\ref{apeE}, all detailed of calculations are presented.


\section{ Model and Method}\label{sec2}
\subsection{General Setup}
\begin{figure}[t] \includegraphics[width=1. \columnwidth]{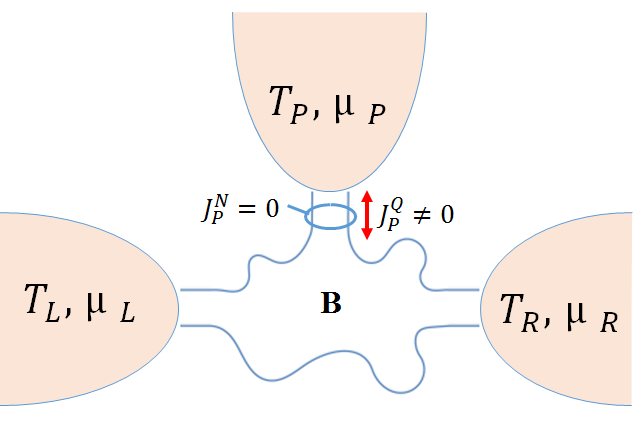}
 \caption{Schematic drawing of a voltage-probe heat engine in the presence of an external magnetic field $(\textbf{B})$.}
\label{f1}
\end{figure}


The voltage-probe setup illustrated in  Fig.~\ref{f1} contains a scattering region in contact with electronic reservoirs with temperatures and chemical potentials, described as $T_\alpha=T+\Delta T_\alpha$ and $\mu_\alpha=\mu+\Delta \mu_\alpha$, where $\alpha=L, P, R$ denotes respective reservoirs. Inelastic scattering phenomena\cite{a443} are simulated using a voltage probe whose temperature and chemical-potential parameters are adjusted to block particle current ($J_P^N=0$) while allowing heat current ($J_P^Q\neq0$). 
We assume the right reservoir ($R$) as a reference so that $\mu_R=\mu$ and $T_R=T$. The setup is also considered to work in the linear response regime  with $|\Delta \mu_\alpha|/k_BT\ll1$ and $|\Delta T_\alpha|/T\ll1$, where $k_{\rm B}$ is the Boltzmann constant. Energy ($J_\alpha^U$ ) and particle ($J_\alpha^N$) currents flowing from the $\alpha_{\rm th}$ reservoir into the scattering region satisfy the current conservation law by the constraint $\sum_\alpha J_\alpha^{U(N)}=0$, where $\alpha=L,R,P$. Heat current $J^Q_\alpha$ is related to the $J_\alpha^U$ and $J_\alpha^N$ by  $J^Q_\alpha=J_\alpha^U-eV_\alpha J_\alpha^N$, where $V_\alpha=\mu_\alpha/e$ is the reservoir voltage. Charge current is directly proportional to the particle current by $J_\alpha^e$=$eJ^N_\alpha$, where $e$ is the electron charge. Note that the positive current values are associated with flows from the respective reservoirs to the scattering region. Using the Landauer-B\"uttiker formalism, it is possible to derive the coherent flow of heat $J_\alpha^Q$ and particles $J_\alpha^N$ via a non-interacting conductor, see appendix.~\ref{apeA} for more details. 
\par
The relation $\dot{\cal S}=\sum_\alpha J^Q_\alpha/T_\alpha$ gives the sum of the entropy production rate, and within the linear response regime, it can be described as $\dot{\cal S}=JX=\sum_{\alpha=1}^4 J_\alpha X_\alpha$, where $J$ and $X$ are four-dimensional vectors defined as follows:
\begin{eqnarray}
J&=&(J_L^N,J_L^Q,J_P^N,J_P^Q),\nonumber\\
X&=&(X_L^{V},X_L^T,X_P^{V},X_P^T),
\label{e1}
\end{eqnarray}
 where $X_\alpha^V$=$\Delta V_\alpha/T$ and $X_\alpha^T$=$\Delta T_\alpha/T^2$ ($\alpha=L, P$) are the generalized forces. Assuming that the thermodynamic forces are small in the linear response regime, the relationship between generalized forces  $X_\alpha$ ( driving irreversible processes) and fluxes $J_\alpha$ (the system's response to the external forces) is linear so that
\begin{eqnarray}
J={\cal L} X,
\label{e2}
\end{eqnarray}
where $\cal L$ is a 4 × 4 Onsager matrix. Eq.~(\ref{e2}) can be recast in a matrix form as follows: 
\begin{equation}
\begin{bmatrix}
J_{L}^N\\
J_{L}^Q\\
J_{p}^N\\
J_{P}^Q
\end{bmatrix}
=
\begin{bmatrix}
{\cal L}_{11} &{\cal L}_{12} & {\cal L}_{13} & {\cal L}_{14}\\
{\cal L}_{21} & {\cal L}_{22} & {\cal L}_{23} & {\cal L}_{24}\\
{\cal L}_{31} & {\cal L}_{32} & {\cal L}_{33} & {\cal L}_{34}\\
{\cal L}_{41} & {\cal L}_{42} & {\cal L}_{43} & {\cal L}_{44}
\end{bmatrix}
\begin{bmatrix}
X_L^{V}\\
X_L^T\\
X_P^{V}\\
X_P^T
\end{bmatrix}
\label{e3}
\end{equation}
where ${\cal L}_{ii}$ and ${\cal L}_{ij}$ are diagonal and off-diagonal elements, respectively, which are given in appendix.~\ref{apeB} (see Eq.~(\ref{eqB1})). Since the probe reservoir neither releases particles nor absorbs ($J_{P}^{N}$=$0$), the problem can reduce to three particle and heat fluxes that are related to the respective generalized forces by the Onsager matrix
$\cal {\cal L}^{\prime}$ as follows: 
\begin{equation}
\begin{bmatrix}
J_L^N\\
J_L^Q\\
J_P^Q\\
\end{bmatrix}
=
\begin{bmatrix}
{\cal L}^{\prime}_{11} &{\cal L}^{\prime}_{12} & {\cal L}^{\prime}_{13} \\
{\cal L}^{\prime}_{21} & {\cal L}^{\prime}_{22} & {\cal L}^{\prime}_{23} & \\
{\cal L}^{\prime}_{31} & {\cal L}^{\prime}_{32} & {\cal L}^{\prime}_{33} & 
\end{bmatrix}
\begin{bmatrix}
X_L^{V}\\
X_L^T\\
X_P^{T}\\
\end{bmatrix}
\label{e7}
\end{equation}
where ${\cal L}^{\prime}_{ij}$ are expressed in appendix.~\ref{apeB} (see Eq.~(\ref{eqB3})).
\par
We notice that the Onsager matrix of the B\"uttiker-probe engine\cite{a45} can be retrieved from Eq.~(\ref{e7}) by imposing $J_P^Q=0$, which leads to the $2\times2$ Onsager matrix as follows:
\begin{equation}
\begin{bmatrix}
J_L^N\\
J_L^Q\\
\end{bmatrix}
=
\begin{bmatrix}
{\cal L}^{\prime\prime}_{11} & {\cal L}^{\prime\prime}_{12} \\
{\cal L}^{\prime\prime}_{21} & {\cal L}^{\prime\prime}_{22} & 
\end{bmatrix}
\begin{bmatrix}
X_L^{V}\\
X_L^T\\
\end{bmatrix}
\label{eq777}
\end{equation}
where ${\cal L}^{\prime\prime}_{ij}$ are given in  appendix.~\ref{apeB} (see Eq.~(\ref{eqB6})). 
\par
Now we expose the setup by an external magnetic field $(\textbf{B})$. With the presence of a magnetic field $(\textbf{B})$, the laws of physics remain constant if time $t$ is replaced by $-t$, provided that the field $\textbf{B}$ is replaced by $-\textbf{B}$. In this case, the Onsager-Casimir relations for the off-diagonal terms of the Onsager matrix are expressed as
\begin{eqnarray}
{\cal L}_{ij}(\textbf{B})={\cal L}_{ji}(-\textbf{B})
\label{e8}
\end{eqnarray}
when $\textbf{B}$ vanishes, the Onsager reciprocal relation
${\cal L}_{ij}={\cal L}_{ji}$ is retrieved.

\subsection{Bounds on Onsager matrix}
For a two-terminal heat engine, the second law of thermodynamics constrains the Onsager matrix coefficients, and it is derived from the positivity of the entropy production rate\cite{a5}, namely ${\dot{\cal S}}\geq0$. Nevertheless, for a three-terminal setup with ${\bold B\neq0}$, current conservation is mathematically expressed by unitary of the scattering matrix, imposing additional bounds on the Onsager matrix elements, more robust than those obtained from the positivity of entropy production rate\cite{a18}. Therefore, it requires to fulfill the following inequalities for a voltage-probe heat engine:
\begin{eqnarray}
\mathfrak{L}_{11}\geq0&,&\mathfrak{L}_{22}\geq0,\nonumber\\
\mathfrak{L}_{11}\mathfrak{L}_{22}&+&\mathfrak{L}_{12}\mathfrak{L}_{21}-[\mathfrak{L}_{12}^2-\mathfrak{L}_{21}^2]\geq0,
\label{e9}
\end{eqnarray}
where
\begin{eqnarray}
\mathfrak{L}_{11}&=&{\cal L}^{\prime}_{11}(\textbf{B}),\nonumber\\
\mathfrak{L}_{12}&=&{\cal L}^{\prime}_{12}(\textbf{B})+{\cal L}^{\prime}_{13}(\textbf{B})\xi,\nonumber\\ 
\mathfrak{L}_{21}&=&{\cal L}^{\prime}_{21}(\textbf{B})+{\cal L}^{\prime}_{31}(\textbf{B})\xi,\nonumber\\
\mathfrak{L}_{22}&=&{\cal L}^{\prime}_{22}(\textbf{B})+{\cal L}^{\prime}_{33}(\textbf{B})\xi^2+\Big[{\cal L}^{\prime}_{23}(\textbf{B})+{\cal L}^{\prime}_{32}(\textbf{B})\Big]{\xi},\nonumber\\
\label{e10}
\end{eqnarray}
and $\xi={X_P^T}/{X_L^T}$. In what follows, we will drop $\textbf{B}$ in the Onsager coefficients for simplicity. By setting the limit, $J_P^Q=0$, the inequalities for a B\"uttiker-probe heat engine can be achieved\cite{a18, a19, a45}, so
\begin{eqnarray}
{\cal L}^{\prime\prime}_{11}\geq0&,&{\cal L}^{\prime\prime}_{22}\geq0,\nonumber\\
{\cal L}^{\prime\prime}_{11}{\cal L}^{\prime\prime}_{22}&+&{\cal L}^{\prime\prime}_{12}{\cal L}^{\prime\prime}_{21}-[{{\cal L}^{\prime\prime}_{12}}^2-{{\cal L}^{\prime\prime}_{21}}^2]\geq0,
\label{e101}
\end{eqnarray}

\subsection{Transport Coefficients}
Onsager coefficients are linked to the transport coefficients, namely, the Seebeck coefficient ($S_{ij}$), electrical conductance $(G_{ij})$, thermal conductance ($K_{ij}$), and Peltier coefficient $(\Pi_{ij})$. Multi-terminal thermoelectric devices introduce non-local transport coefficients with index ${i\neq j}$, describing how applied bias between two terminals influences transport properties in the remaining terminals\cite{a201}.
\par
For a multi-terminal heat engine, $S_{ij}$ is given by the relationship between potential and temperature biases between terminals, assuming no net particle current is flowing through the system\cite{a201}
\begin{eqnarray}
 {S_{ij}=-\frac{\Delta V_i}{\Delta T_j}}_{\binom{ {J_\alpha^N=0~\forall \alpha;}}{ \Delta T_\alpha=0~\forall {\alpha\ne j}}},
 \label{e11}
\end{eqnarray}
and for a voltage-probe heat engine, it can be described as follows:
\begin{eqnarray}
S_{LL}(\textbf{B})=\dfrac{{\cal L}^{\prime}_{12}}{T{\cal L}^{\prime}_{11}} &,& S_{LP}(\textbf{B})=\dfrac{{\cal L}^{\prime}_{13}}{T{\cal L}^{\prime}_{11}},
\label{e12}
\end{eqnarray}
\begin{eqnarray}
S_{LL}(-\textbf{B})=\dfrac{{\cal L}^{\prime}_{21}}{T{\cal L}^{\prime}_{11}} &,& S_{LP}(-\textbf{B})=\dfrac{{\cal L}^{\prime}_{31}}{T{\cal L}^{\prime}_{11}},
\label{e13}
\end{eqnarray}
where $S_{LL}$ and  $S_{LP}$ represent local and non-local coefficients, respectively. Using Eq.~(\ref{eq777}), we can derive the Seebeck coefficient for the B\"uttiker-probe heat engine as follows:
\begin{eqnarray}
S_{LL}(\textbf{B})=\dfrac{{\cal L}^{\prime\prime}_{12}}{T{{\cal L}^{\prime\prime}_{11}}} &,& S_{LL}(-\textbf{B})=\dfrac{{\cal L}^{\prime\prime}_{21}}{T{{\cal L}^{\prime\prime}_{11}}}
\label{eqC7}
\end{eqnarray}
 The electrical conductance for the multi-terminal setup  generalizes to\cite{a201}
\begin{eqnarray}
 {G_{ij}=\frac{J_i^N}{\Delta V_j}}_{\binom{ { \Delta V_\alpha=0~\forall {\alpha\ne j};}}{ \Delta T_\alpha=0~\forall \alpha}},
\label{e14}
\end{eqnarray}
and for a voltage-probe heat engine, it is given as
\begin{eqnarray}
 G_{LL}=\dfrac{{\cal L}^{\prime}_{11}}{T},
\label{e15}
\end{eqnarray}
where $G_{LL}$ denotes the local electrical coefficient. For the B\"uttiker-probe engine,  it can be described as
\begin{eqnarray}
 G_{LL}=\dfrac{{\cal L}^{\prime\prime}_{11}}{T}.
\label{eqC8}
\end{eqnarray}
The thermal conductance is expressed as\cite{a201} 
\begin{eqnarray}
 {K}_{ij}=\frac{J_i^Q}{\Delta T_j}_ {\binom{{J_\alpha^N=0~\forall \alpha;}}{ \Delta T_\alpha=0~\forall {\alpha\ne j}}},
 \label{e16}
\end{eqnarray}
and for a voltage-probe engine, it can be written in terms of the Onsager coefficients with local $K_{ii}$ and non-local $K_{ij}$ coefficients as
\begin{eqnarray}
{ K}_{LL}=\dfrac{{\cal L}^{\prime}_{22}{\cal L}^{\prime}_{11}-{\cal L}^{\prime}_{21}{\cal L}^{\prime}_{12}}{T^2{\cal L}^{\prime}_{11}}&,&
 {K}_{PP}=\dfrac{{\cal L}^{\prime}_{33}{\cal L}^{\prime}_{11}-{\cal L}^{\prime}_{31}{\cal L}^{\prime}_{13}}{T^2{\cal L}^{\prime}_{11}},\nonumber\\
 {K}_{LP}=\dfrac{{\cal L}^{\prime}_{11}{\cal L}^{\prime}_{23}-{\cal L}^{\prime}_{21}{\cal L}^{\prime}_{13}}{T^2{\cal L}^{\prime}_{11}}&,&
 {K}_{PL}=\dfrac{{\cal L}^{\prime}_{11}{\cal L}^{\prime}_{32}-{\cal L}^{\prime}_{12}{\cal L}^{\prime}_{31}}{T^2{\cal L}^{\prime}_{11}} ,\nonumber\\
\label{e17}
\end{eqnarray}
and for the B\"uttiker-probe heat engine, it gives
\begin{eqnarray}
K_{LL}=\dfrac{{\cal L}^{\prime\prime}_{22}{\cal L}^{\prime\prime}_{11}-{\cal L}^{\prime\prime}_{21}{\cal L}^{\prime\prime}_{12}}{T^2{\cal L}^{\prime\prime}_{11}}.
\label{eqC9}
\end{eqnarray}
In any thermoelectric device, the Peltier coefficient is related
to the Seebeck coefficient as $\Pi _{ij}(\textbf{B})$=$TS_{ji}(-\textbf{B})$ when $\textbf{B}\ne 0$.
\subsection{Thermodynamic Efficiency of the Steady-State Heat Engines}
The efficiency of any steady-state heat engine is bounded from above by the Carnot-engine efficiency $\eta_c$, explained for a three-terminal heat engine in appendix.~\ref{apeC}. Given that a voltage-probe quantum heat engine is operating between three reservoirs at different temperatures $T_L$, $T_P$, and $T_R$, the efficiency is defined as\cite{a201}
\begin{eqnarray}
\eta&=&\frac{\cal P}{\sum_{\alpha+}J^Q_\alpha}\leq \eta_c\nonumber\\
&=&\frac{J^Q_L+J^Q_P+J^Q_R}{\sum_{\alpha+}J^Q_\alpha}=\frac{-TJ^N_LX_L^V}{\sum_{\alpha+}J^Q_\alpha}
\label{e19}
\end{eqnarray}
where ${\cal P}>0$ is the output power of the heat engine, equalling the sum of all heat exchanged between the scattering region and reservoirs, and the symbol $\sum_{\alpha+}$ in the denominator is only restricted to positive heat currents ($J^Q_\alpha>0$, where $\alpha=L, P$). For the voltage-probe quantum heat engine sketched in Fig.~\ref{f1}, for simplicity, we set $T_L>T_P>T_R$ and consider only those situations where  $J^{Q}_R$ is supplied by the scattering region ($J^Q_R<0$). It is worth mentioning that within the regime that $J_R^Q>0$, the system effectively works like a refrigerator, absorbing heat from the coldest reservoir(s). Taking these assumptions, when both $J^Q_L$ and $J^Q_P$ are absorbed from  the respected reservoirs, the efficiency reads
\begin{eqnarray}
\eta_{LP}=\frac{\cal P}{J^Q_L+J^Q_P},
\label{e20}
\end{eqnarray}
and when either $J^Q_L$ or $J^Q_P$ is extracted from the respected reservoir, efficiency reads
\begin{eqnarray}
\eta_{L,(P)}=\frac{\cal P}{J^Q_{L,(P)}},
\label{e21}
\end{eqnarray}
the subscript $L$ in the denominator is replaced by $P$ if the probe reservoir is the only one that releases heat into the scattering region and two other reservoirs absorb heat from it. Since the signs of the heat currents flowing throughout the scattering region and reservoirs are not a priory; in other words, it depends upon the details of the system, the efficiency expression relies on the heat current(s), injected into the scattering region from the respective reservoir(s). Note that, in the formalism developed in the following we refer to the efficiency of the voltage probe as $\eta_m$, with sub-index $m=L,P,LP$. This discriminates it from the efficiency of  B\"uttiker probe we refer to it with $\eta$.
\subsection{Efficiency at Maximum Output Power}
Here we formulate $\eta_m(\mathcal{P}_{\rm max})$ in the context of the irreversible heat engine,  for the three terminal voltage-probe setup. From Eq.~(\ref{e7}), it can be found that the output power of a voltage-probe heat engine is a function of three generalized forces, namely $X_L^V$, $X_L^T$ and $X_P^T$, and it can be written as
\begin{eqnarray}
\mathcal{P}=-T(J_{L}^{N}X_{L}^{V})>0,
\label{e22}
\end{eqnarray}
where $J_L^N={\cal L}^{\prime}_{11}X_L^V+\big({\cal L}^{\prime}_{12}+{\cal L}^{\prime}_{13}\xi\big)X_L^T$. To derive the maximum output power of the heat engine, the derivative of $\mathcal{P}$ with respect to $X_L^V$  is calculated while $X_L^T$ and $X_P^T$ are kept constant, so 
\begin{eqnarray}
{{X}_{L}^{V}}^{*}=\frac{-\mathfrak{L}_{12}}{2\mathfrak{L}_{11}}X_{L}^{T}.
\label{e23}
\end{eqnarray}
Inserting ${X_L^V}^*$ into Eq.~(\ref{e22}), the maximum output power can be defined as
\begin{eqnarray}
\mathcal{P}_{\rm max}=\frac{T^4}{4}\mathbb{G}\mathbb{S}^2 {X_L^T}^2,
\label{e24}
\end{eqnarray}
where $\mathbb{G}=G_{LL}$ and $\mathbb{S}=(S_{LL}+S_{LP}\xi)$.
Exploiting Eqs.~(\ref{e23}) and (\ref{e24}), efficiency at maximum output power  for different  cases detailed in Eqs.~(\ref{e20}) and (\ref{e21}) can be derived as follows:
\begin{eqnarray}
\eta_{m}(\mathcal{P}_{\rm max})=\dfrac{\eta_{c,{m}}(\mathcal{P}_{\rm max})}{2}\dfrac{x_my_m}{y_m+2d_m}&;& m=L, P, LP.\nonumber\\
\label{e241}
\end{eqnarray}
In appendix.~\ref{apeD}, we present the details of derivation for Eq.~(\ref{e241}). In Eq.~(\ref{e241}), $\eta_{c,m}(\mathcal{P}_{\rm max})$ is the value of the Carnot-engine efficiency which is derived by setting $X_L^V={X_L^V}^*$ in Eqs.~(\ref{eqA12}) and (\ref{eqA13}), and $x_m$ is the asymmetry parameter that is expressed as
\begin{eqnarray}
x_m=\frac{r_m}{y_m},
\label{e25}
\end{eqnarray}
where
\begin{eqnarray}
r_m=\big(2\delta{\cal Z}_m^A+{\cal Z}_m^B+\delta^2{\cal Z}_m^C\big)T,
\label{e2551}
\end{eqnarray}
and $y_m$ is the generalized figure of merit, which is described as follows:
\begin{eqnarray}
y_m=\Big({\delta \big({\cal Z}_m^{A^\prime}+{\cal Z}_m^{A^{\prime\prime}}\big)+{\cal Z}_m^{B^\prime}+{\delta}^2 {\cal Z}_m^{C^\prime}}\Big)T, 
 \label{e26}
\end{eqnarray}
where $\delta=1/\xi$. The terms ${\cal Z}_m^\theta T$ with the superscripts $\theta=A, A^\prime, A^{\prime\prime}, B, B^\prime, C$ and $C^\prime$, are given in Eq.~(\ref{eqd2}). The parameter $d_m$ in Eq.~(\ref{e241}) are given as follows:
\begin{eqnarray}
d_L=\delta\frac{K_{PL}+K_{LP}}{K_{LL}}+\frac{K_{PP}}{K_{LL}}+{\delta}^2 &~~{\rm if}~~& J^Q_L>0,\nonumber\\
d_P=\delta\frac{K_{PL}+K_{LP}}{K_{PP}}+{\delta}^2\frac{K_{LL}}{K_{PP}}+1&~~{\rm if}~~& J^Q_P>0, \nonumber\\   
d_{LP}=\frac{\delta{K}_{PL}+K_{PP}}{K_{LP}}+{\delta}^2\frac{K_{LL}}{K_{LP}}+\delta&~~{\rm if}~~& J^Q_L,J^Q_P>0.\nonumber\\
\label{e27}
\end{eqnarray}
The parameter $d_m$ is related to the model's thermal conductance and the temperature bias ratio. Utilizing a toy model, as illustrated in the appendix.~\ref{apeE}, we show the possible range of values for $d_m$. Furthermore, $d_m$ represents the main difference between the voltage-probe and the B\"uttiker-probe, since our setup includes more complexity and potential for fine-tuning to improve efficiency.

For the time-symmetric case ($x_m=1$), Eq.~(\ref{e26}) reduces to 
\begin{eqnarray}
{\mathbb{Z}}_mT=\big(2\delta{\cal Z}_m^{A_1}+{\cal Z}_m^{B_1}+\delta^2{\cal Z}_m^{C_1}\big)T,
\label{e2626}
\end{eqnarray}
 where $A_1=S_{LL}S_{LP}G_{LL}$, $B_1=S_{LP}^{2}G_{LL}$ and $C_1=S_{LL}^{2}G_{LL}$. We noticed that Eq.~(\ref{e2626}) is the more compact version of the general figure of merit, which is derived by Mazza et al.\cite{a201} for a three-terminal genuine heat engine within time-reversal symmetry case.

As discussed below, the formalism developed for the voltage-probe setup can achieve the B\"uttiker-probe model by imposing the limit of $J_P^Q=0$. When $J_P^Q=0$, Eq.~(\ref{e2551}) reduces to
\begin{eqnarray}
r_L=\delta^2{\cal Z}_L^CT,
\label{e2611}
\end{eqnarray}
where 
\begin{eqnarray}
{\cal Z}_L^CT=\dfrac{S_{LL}(B)^2G_{LL}(B)T}{K_{LL}},
\label{e2612}
\end{eqnarray}
and correspondingly Eq.~(\ref{e26}) reduces to
\begin{eqnarray}
y_L={\delta}^2 {\cal Z}_L^{C^{\prime}}T,
\label{e261}
\end{eqnarray}
where 
\begin{eqnarray}
{\cal Z}_L^{C^{\prime}}T=y=\dfrac{G_{LL}S_{LL}(\textbf{B})S_{LL}(-\textbf{B})T}{K_{LL}},
\label{e262}
\end{eqnarray}
is the general figure of merit of the B\"uttiker-probe heat engine with broken time-reversal symmetry\cite{a18, a19, a45}.

By means of Eqs.~(\ref{e2611}) and (\ref{e261}), Eq.~(\ref{e25}) can be rewritten for the B\"uttiker-probe engine as follows:
\begin{eqnarray}
{\textit{x}}=\dfrac{S_{LL}(\textbf{B})}{S_{LL}(-\textbf{B})}.
\label{e251}
\end{eqnarray}
Finally, in the case that $J_P^Q$=0, Eq.~(\ref{e27}) reduces to
\begin{eqnarray}
d_L={\delta}^2. 
\label{e271}
\end{eqnarray}
Inserting Eqs.~(\ref{e261}), (\ref{e251}) and (\ref{e271}) into Eq.~(\ref{e241}), the efficiency at maximum output power can be expressed for the B\"uttiker-probe heat engine as follows:
\begin{eqnarray}
\eta({\cal P}_{\rm max})=\dfrac{\eta_{c}}{2}\dfrac{{\textit{x}}{y}}{{y+2}},
\label{e242}
\end{eqnarray}
where $\eta_c=\frac{{\Delta T}_L}{T}$ is the Carnot efficiency of a B\"uttiker-probe engine. For $x=1$, the conventional figure of merit for the B\"uttiker probe setup can be expressed as
 \begin{eqnarray}
 {\mathbb{Z}}T=\frac{G_{LL}S_{LL}^{2}T}{K_{LL}}. 
\label{e2625}
\end{eqnarray}
 and the maximum output power efficiency for the time-symmetric case is given by $\eta({\cal P}_{\rm max})=\frac{\eta_{c}}{2}\frac{{\mathbb{Z}}T}{{\mathbb{Z}T}+2}$.

 \subsection{Efficiency at a Given Output Power}
To examine the thermoelectric properties of a three-terminal heat engine at a given output power, the relative gains in power $\Delta {\cal P}$ can be expressed as follows:
\begin{eqnarray}
\Delta {\cal P}&=&\frac{ {\cal P}-{\cal P}_{\rm max}}{{\cal P}_{\rm max}}.
\label{e28}
\end{eqnarray}

 The power gain, $\Delta {\cal P}$, provides us with precise results on the thermoelectric performance of heat engines at a given output power. After simplicity
\begin{eqnarray}
\frac{{\cal P}}{{\cal P}_{\rm max}}=1+\Delta{\cal P}=\frac{J_L^NX_L^V}{{J_L^N}^*{X_L^V}^{*}}=\varepsilon(2-\varepsilon),
\label{e29}
\end{eqnarray}
where $\varepsilon=X_L^V/{X_L^V}^*$. The normalized efficiency can be expressed as follows:
\begin{eqnarray}
\frac{\eta_m}{\eta_m({\cal P}_{\rm max})}=\varepsilon(2-\varepsilon)\dfrac{y_m+2d_m}{2\Big(y_m+d_m\Big)-y_m\varepsilon}.
\label{e291}
\end{eqnarray}
Using  Eqs.~(\ref{e261}) and (\ref{e271}), one can derive the normalized efficiency for a B\"uttiker-probe engine\cite{a45} as follows:
\begin{eqnarray}
\frac{\eta}{\eta({\cal P}_{\rm max})}=\varepsilon(2-\varepsilon)\dfrac{{{y}}+2}{2\Big({{y}}+1\Big)-{{y}}\varepsilon}.
\label{e2911}
\end{eqnarray}
Using Eq.(\ref{e29}), efficiency dependence on the output power is defined as
\begin{eqnarray}
\varepsilon_{\pm}=1\pm\sqrt{-\Delta {\cal P}}.
\label{e31}
\end{eqnarray}
The plus sign, here, is related to the favorable case where the normalized efficiency goes beyond the ${\cal P}/{\cal P}_{\rm max}$ as the external force is increased,
and the minus sign describes the opposite case. Furthermore, the ratio $\eta_m/\eta_m({\cal P}_{\rm max})$ can be expressed as
\begin{eqnarray}
\frac{\eta_m}{\eta_m({\cal P}_{\rm max})}=(1+\Delta{\cal P})\dfrac{y_m+2d_m}{2(y_m+d_m)-y_m(1\pm \sqrt{-\Delta{\cal P}})},\nonumber\\
\label{e32}
\end{eqnarray}
and for the B\"uttiker-probe engine\cite{a45}, Eq.~(\ref{e32}) can be rewritten as
\begin{eqnarray}
\frac{\eta}{\eta({\cal P}_{\rm max})}=(1+\Delta{\cal P})\dfrac{{{y}}+2}{2({{y}}+1)-{{y}}(1\pm \sqrt{-\Delta{\cal P}})}.
\label{e321}
\end{eqnarray}
Moreover, the efficiency $\eta_m$ at a given output power as a function of the power gain $\Delta{\cal P}$ can be derived as
\begin{eqnarray}
\eta_m=\frac{\eta_{c,m}}{2}\dfrac{x_my_m(1+\Delta{\cal P})}{2(y_m+d_m)-(1\pm \sqrt{-\Delta{\cal P}})y_m},
\label{e33}
\end{eqnarray}
and for the B\"uttiker-probe engine\cite{a45}, Eq.(\ref{e33}) can be expressed as
\begin{eqnarray}
\eta=\frac{\eta_{c}}{2}\dfrac{\textit{x}y(1+\Delta{\cal P})}{2({{y}}+1)-(1\pm \sqrt{-\Delta{\cal P}})y}.
\label{e331}
\end{eqnarray}
\begin{figure}[t]
\includegraphics[width=1\columnwidth]{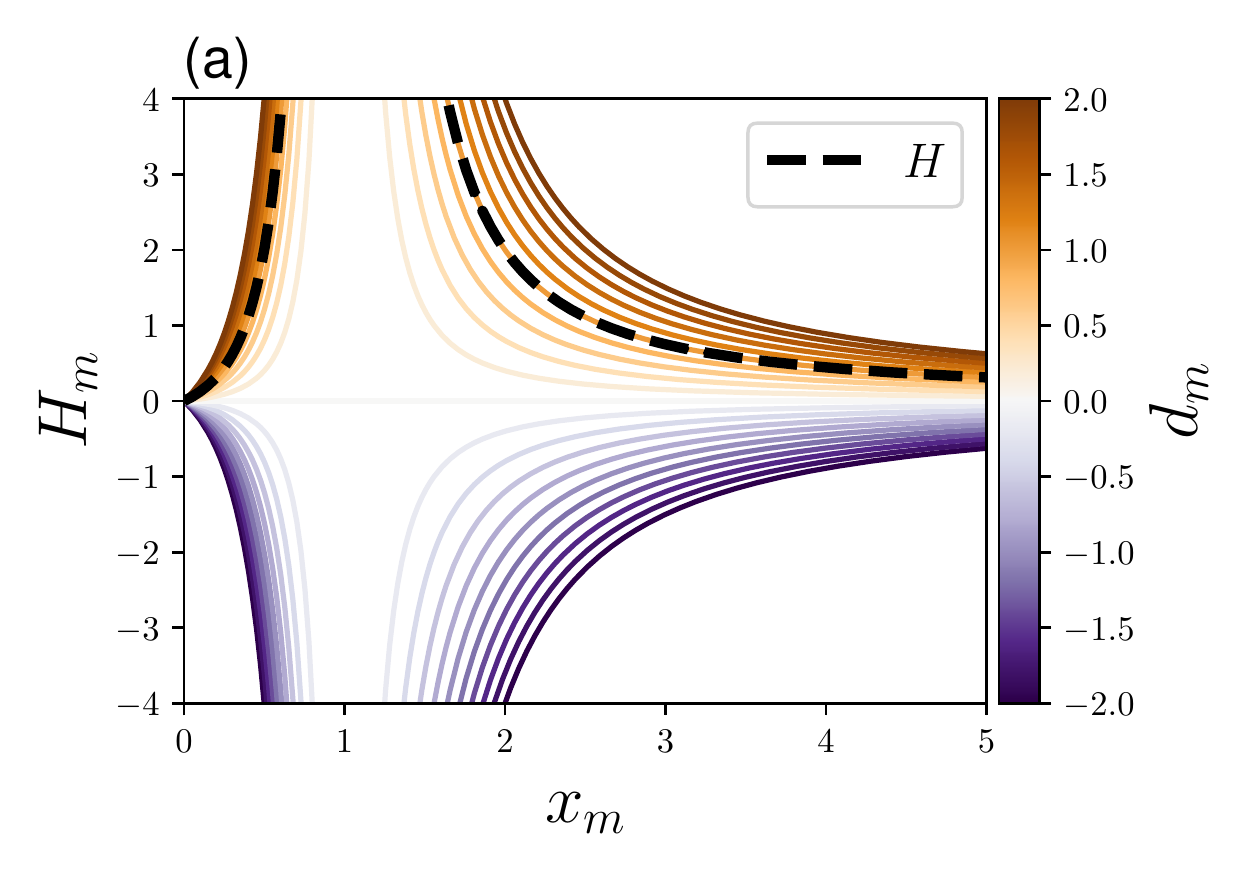}
\includegraphics[width=1\columnwidth]{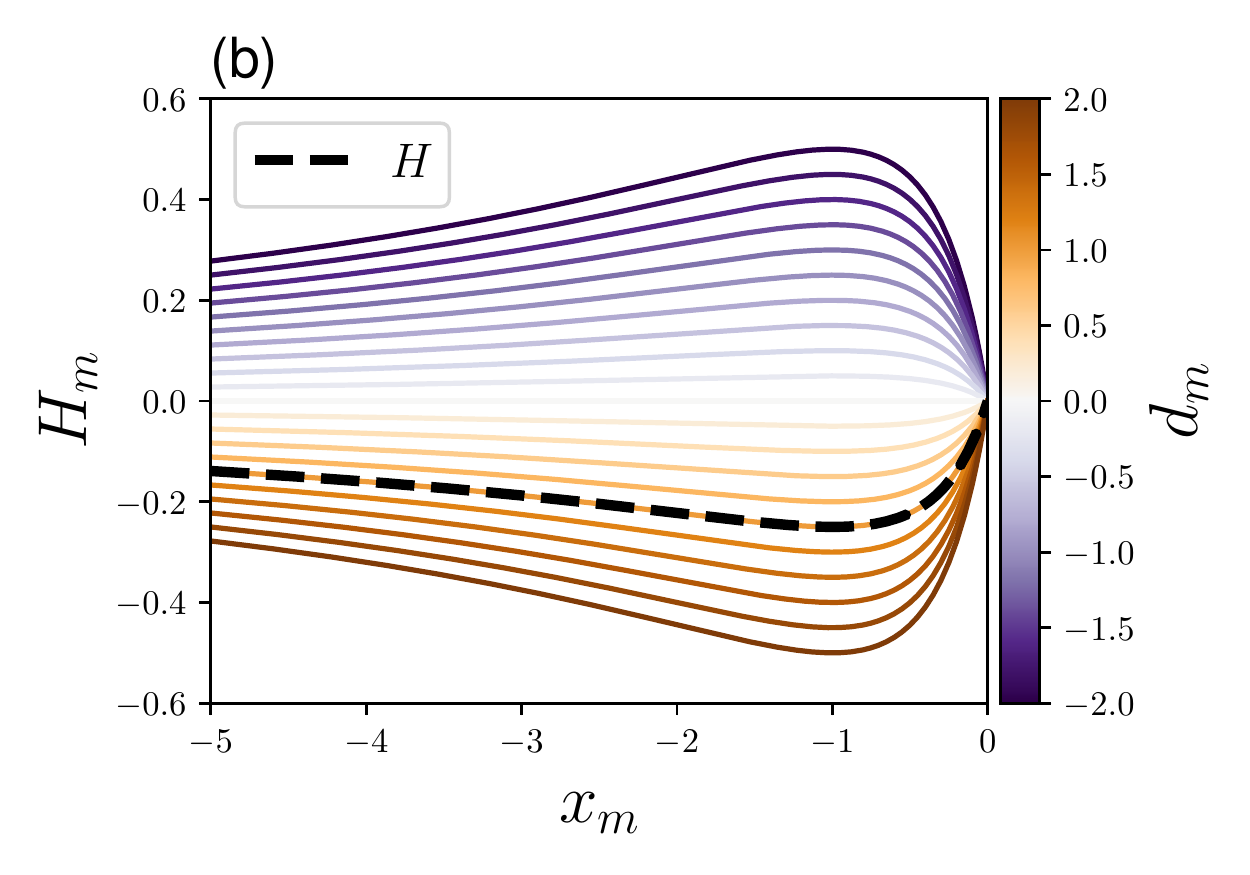}
\caption{ The function $H_m$ versus $x_m$ for different $d_m$ values are color-coded. Dashed-black lines display $H$ (Eq.~\ref{e37}).}
\label{f2}
\end{figure}
\begin{figure*}[t]
\includegraphics[width=2\columnwidth]{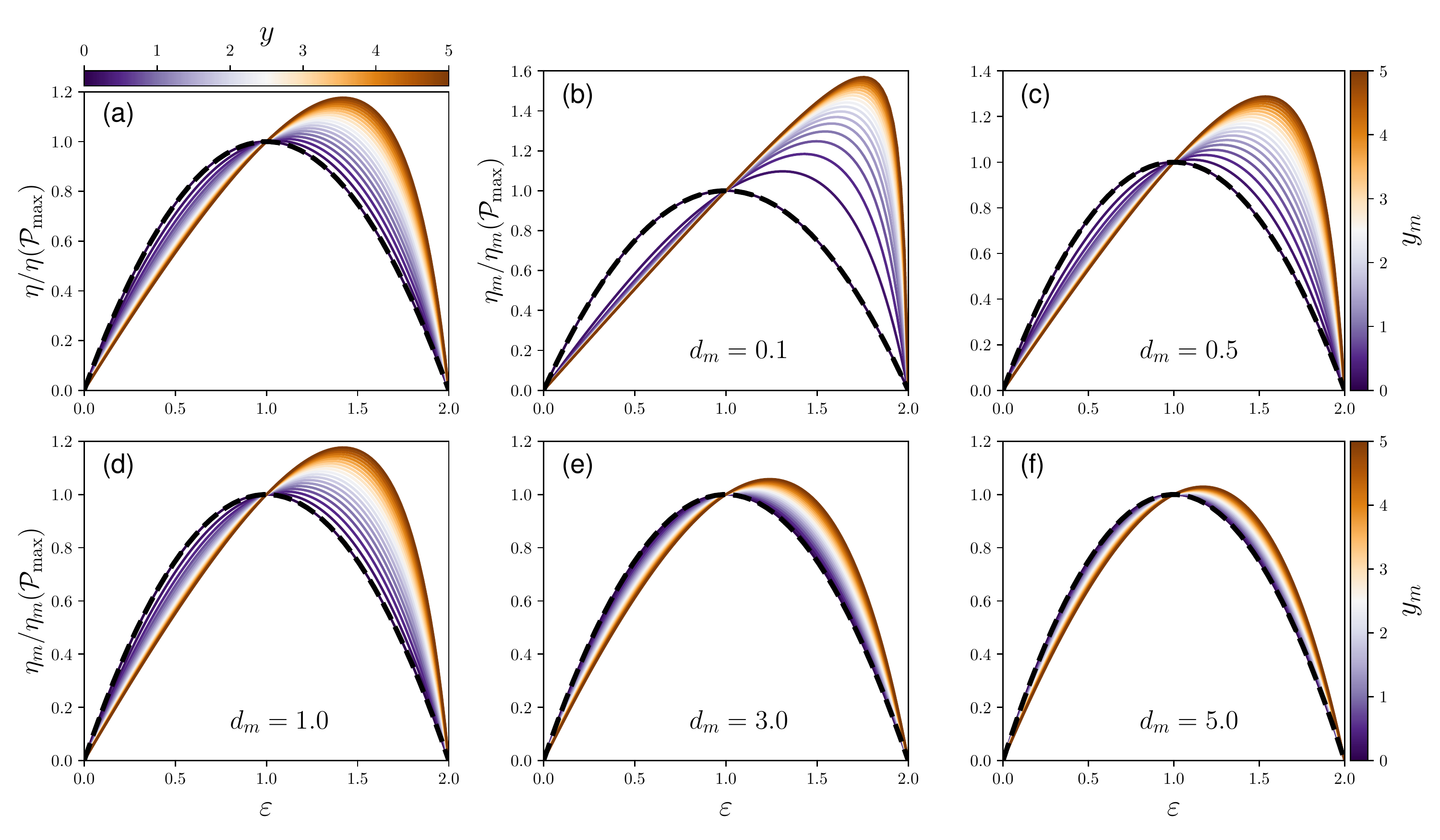}
\caption{Ratio $\eta_m/\eta_m({\cal P}_{\rm max})$ as a function of $\varepsilon$ for different values of $y_m$ and $d_m$ in the region of $x_m>0$. Dashed-black lines show ${\cal P}/{{\cal P}_{\rm max}}$. (a) demonstrates the normalized efficiency of a B\"uttiker-probe heat engine as a function of $\varepsilon$ for different values of $y$-(b), (c), (d), (e), an (f) show the normalized efficiency of a voltage-probe heat engine for $d_m=0.1,~0.5,~1.0,~3.0,~5.0$, respectively, and  different values of $y_m$ are color coded.}
\label{fig3}
\end{figure*}
\begin{figure*}[t]
\includegraphics[width=2\columnwidth]{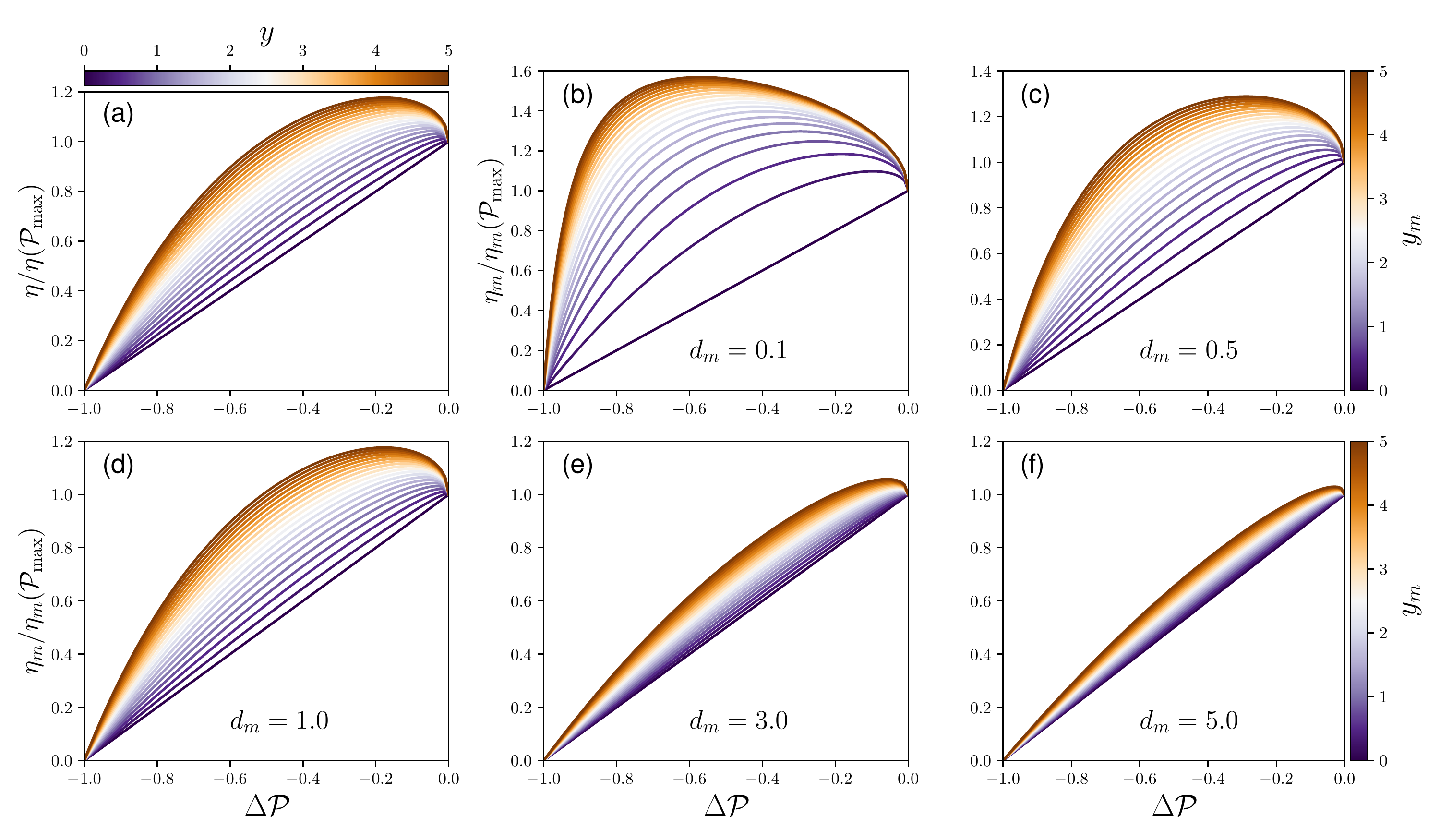}
\caption{The normalized efficiency as a function of power gains $\Delta {\cal P}$ for the favorable case $\varepsilon_+$. The rest parameters are set the same  as the Fig.~\ref{fig3}.}
\label{fig4}
\end{figure*}
\section{Results and Discussion}\label{sec3}
In this section, we provide numerical results of the theory developed in the previous section. We demonstrate how voltage-probe heat engines in the presence of an external magnetic field can produce higher thermoelectric performance than B\"uttiker-probe heat engines\cite{a18, a19, a45}. Following that, we uncover similarities and differences in efficiency optimization between the two setups for a given output power.
\par
From Eq.~(\ref{e331}), it can be seen that the efficiency of a B\"uttiker-probe heat engine relies on the parameters $x$ and $y$. However, the efficiency of a voltage-probe heat engine, in addition to $x_m$ and $y_m$,  depends on the characteristic parameter $d_m$ (see Eq.~(\ref{e33})). Similar to the work by Brandner et al.\cite{a18}, let us first introduce the following inequality for a voltage-probe heat engine with broken time-reversal symmetry using Eqs.~(\ref{e9}), (\ref{e25}) and (\ref{e26}) as follows:
\begin{eqnarray}
H_m\ll y_m\ll0 &~~{\rm for}~~& x_m<0,\nonumber\\
H_m\gg y_m\gg0 &~~{\rm for}~~& x_m>0,
\label{e34}
\end{eqnarray}
where
\begin{eqnarray}
H_m=\frac{d_mx_m}{(x_m-1)^2}.
\label{e35}
\end{eqnarray}
Substituting Eqs.~(\ref{e261}) and (\ref{e271}) into the inequality of Eq.~(\ref{e34}), the inequality of B\"uttiker-probe heat engine within broken time-reversal symmetry achieves as follows:
\begin{eqnarray}
H\ll y\ll 0 &~~{\rm for}~~& x<0, \nonumber\\
H\gg y\gg 0 &~~{\rm for}~~& x>0,
\label{e36}
\end{eqnarray}
where $y$ and $x$ are given in Eq.~(\ref{e262}) and Eq.~(\ref{e251}), respectively, and the function $H$ is given as
\begin{eqnarray}
H=\frac{x}{(x-1)^2}.
\label{e37}
\end{eqnarray}
\par
Fig.~(\ref{f2}) illustrates $H_m$ versus $x_m$ for different $d_m$ values. In appendix.~\ref{apeE}, we display the contour plots of $d_m$ for a triple quantum dot attached to three electronic reservoirs. The results reveal that $d_L$ and $d_P$ are positive, while $d_{LP}$ is negative. Therefore, the positive values of $d_m$ are associated with $m=L, P$, while the negative ones correspond to $m=LP$. In contrast with a B\"uttiker-probe heat engine, where the function $H$ relies only on $x$, for a voltage-probe setup, $H_m$ also depends upon $d_m$, which based on $d_m$ sign and value, bounds on $y_m$ can change. From Fig.~(\ref{f2}), It can been that for positive $d_{m}$, $y_{m}$ is located in $[-0.25, 0]$  and $[0,\infty]$ for $x_{m}<0$ and $x_{m}>0$, respectively. However, for negative $d_{m}$, $y_{m}$ shows a reversed behavior, so that it is located in $[0,0.25]$ and $[-\infty,0]$ for $x_{m}<0$ and $x_{m}>0$, respectively. We noticed that when $x_m=1$, ${\mathbb{Z}}_{L(P)}\rightarrow +\infty$ and ${\mathbb{Z}}_{LP} \rightarrow -\infty$ (see Eq.~(\ref{e2626})), corresponding to the work of Mazza et al.\cite{a201} When $d_m=1$, the function $H_m$ coincides with $H$, as shown by dashed-black lines, and in the case that $x=1$, $\mathbb{Z}T\rightarrow+\infty$ (see Eq.~(\ref{e2625})).
\par
Eq.~(\ref{e29}) shows that the ratio ${\cal P}/{\cal P}_{\rm max}$ is a parabolic curve that depends only on the external load $\varepsilon$. The normalized efficiency of a B\"uttiker-probe heat engine $\eta/\eta({\cal P}_{\rm max})$, expressed in Eq.~(\ref{e2911}), relies on $\varepsilon$ and $y$. However, the normalized efficiency of a voltage-probe engine $\eta_m/\eta_m({\cal P}_{\rm max})$, besides $\varepsilon$ and $y_m$, depends on the additional parameter $d_m$ (see Eq.(\ref{e291})). To determine the efficiency dependence on $d_m$, in Fig.~(\ref{fig3}), we plot the normalized efficiency of the heat engines as a function of $\varepsilon$ when $x_m>0$. We also plot ${\cal P}/{\cal P}_{max}$, shown by dashed-black lines. Fig.~\ref{fig3}.(a) displays $\eta/\eta({\cal P}_{\rm max})$ versus $\varepsilon$ for $y>0$, and Figs.~\ref{fig3}.(b)-(f) show  the behaviors of $\eta_m/\eta_m({\cal P}_{\rm max})$ for different $d_m$ values, as a function of $\varepsilon$  with $y_m$ which is constrained by $x_m>0$. Note that as depicted in Figs.~\ref{f2}.(a,b), when $x_m>0$, $y_{L(P)}$ is located in $[0,\infty]$ for different $d_{L(P)}$, while $y_{LP}$ is located in $[-\infty, 0]$ for different $d_{LP}$. Therefore, in the region of $x_m>0$, we set the same signs for $d_m$ and $y_m$. For the B\"uttiker-probe heat engine, $y$ is located in $[0,\infty]$ (see dashed-black line in Fig.~\ref{f2}.(a)). 

As illustrated in Fig.~\ref{fig3}, for lower external loads $\varepsilon<1$, B\"uttiker and voltage-probe engines show similar characteristics so that the efficiency increases as the output power enhances while it demonstrates a decreasing feature as $y$ and $y_m$ increase. In contrast, for higher external loads $\varepsilon>1$, the efficiency exhibits different behavior; the optimum efficiency value is achieved for $\varepsilon>1$, and it is gone up with increasing the absolute value of figures of merit. 
\par
Let us now elaborate on  how a voltage-probe heat engine with broken time-reversal symmetry can enhance thermoelectric properties at a given output power with respect to a B\"uttiker-probe setup. From Figs.~\ref{fig3}-(b, c), it can be seen that when $d_m<1.0$, for identical values of $y$  and $y_m$, the optimum values of efficiency within $\varepsilon>1$ are larger than those of the B\"uttiker-probe heat engine (see Fig.~\ref{fig3}.(a)), implying that for the voltage-probe heat engine, larger efficiency can be achievable at lower $y_m$  compared to $y$ for a B\"uttiker-probe heat engine. This is due to the presence of the parameter $d_m$ in a voltage-probe setup, contributing to the heat dissipation rate\cite{a443}, $\dot{Q} = T\dot{\cal S}$. For the voltage-probe setup when $d_m<1.0$, $\dot{Q}$ decreases more compared to that of the B\"uttiker-probe heat engine, giving rise to the larger optimum values of efficiency for the voltage probe heat engine in the region of $\varepsilon>1$. However, the opposite occurs when $d_m>1.0$, so the optimum values of efficiency diminish compared with that of the B\"uttiker-probe engine, as shown in Figs.~\ref{fig3}-(e,f). Furthermore, in the case that $d_m=1.0$, the normalized efficiency of the voltage-probe setup exactly shows the same behavior as that of a B\"uttiker-probe heat engine (compare Figs.~\ref{fig3}-(a) and (d)). Therefore, It is interesting that for a voltage-probe heat engine, the parameter $d_m$ is significantly contributing to changing the normalized-efficiency values at a given output power.
\par
 Eq.~(\ref{e31}) indicates that $\varepsilon$ as a function of power gain ($\Delta{\cal P}$) has two branches. To examine the efficiency dependence on $\Delta{\cal P}$, in Fig.~\ref{fig4}, we plot the normalized efficiency as a function of $\Delta{\cal P}$ for the favorable case  $\varepsilon_{+}$. Fig.~\ref{fig4}.(a) exhibits the normalized efficiency of the B\"uttiker-probe heat engine  as a function of $\Delta{\cal P}$ (see Eq.~(\ref{e321})). We display the normalized efficiency of the voltage-probe engines as a function of $\Delta{\cal P}$ for different $d_m$ as shown in Figs.~\ref{fig4}.(b)-(f). 

 From the color-coded lines in Fig.~\ref{fig4}, one can observe that increasing $y_m$($y$) for a fixed $\Delta{\cal P}$ results in enhanced efficiency of a voltage-probe (B\"uttiker-probe) heat engine. Fig.~\ref{fig4}(a) indicates that for a B\"uttiker-probe heat engine, most normalized-efficiency curves monotonically rise as the output power increases (from $-1$ to zero). However, for the voltage-probe engines, when $d_m<1.0$, the normalized efficiency curves mostly non-monotonically change as power gains change, as shown in Figs.~\ref{fig4} (b) and (c). Thus, in the case that $d_m<1.0$, the voltage-probe heat engine achieves a larger efficiency value at lower values of $y_m$ compared to a B\"uttiker probe. When $d_m>1.0$, the normalized efficiency curves monotonically increase as $\Delta{\cal P}$ changes, as depicted in Figs.~\ref{fig4} (e) and (f). Therefore, the voltage-probe engine cannot enhance the optimum value of efficiency, even if $y_m$ increases. For some values of $y_m$ in Fig.~\ref{fig4}, it can be noticed that a larger efficiency can be achieved when the output power ${\cal P}$ is less than the maximum output power ${\cal P}_{\rm max}$. This implies that there is a fundamental trade-off between the efficiency and output power in thermoelectric performance optimization of both B\"uttiker and voltage-probe engines.
In contrast with a B\"uttiker-probe heat engine, whose efficiency depends on $y$ and $\Delta{\cal P}$, the normalized efficiency of the voltage-probe engine indicates that in practice, the operational conditions of the voltage-probe engines are optimized based on $d_m$, $\Delta{\cal P}$ and $y_m$.
\par
\begin{figure}[t]
\includegraphics[width=1.\columnwidth]{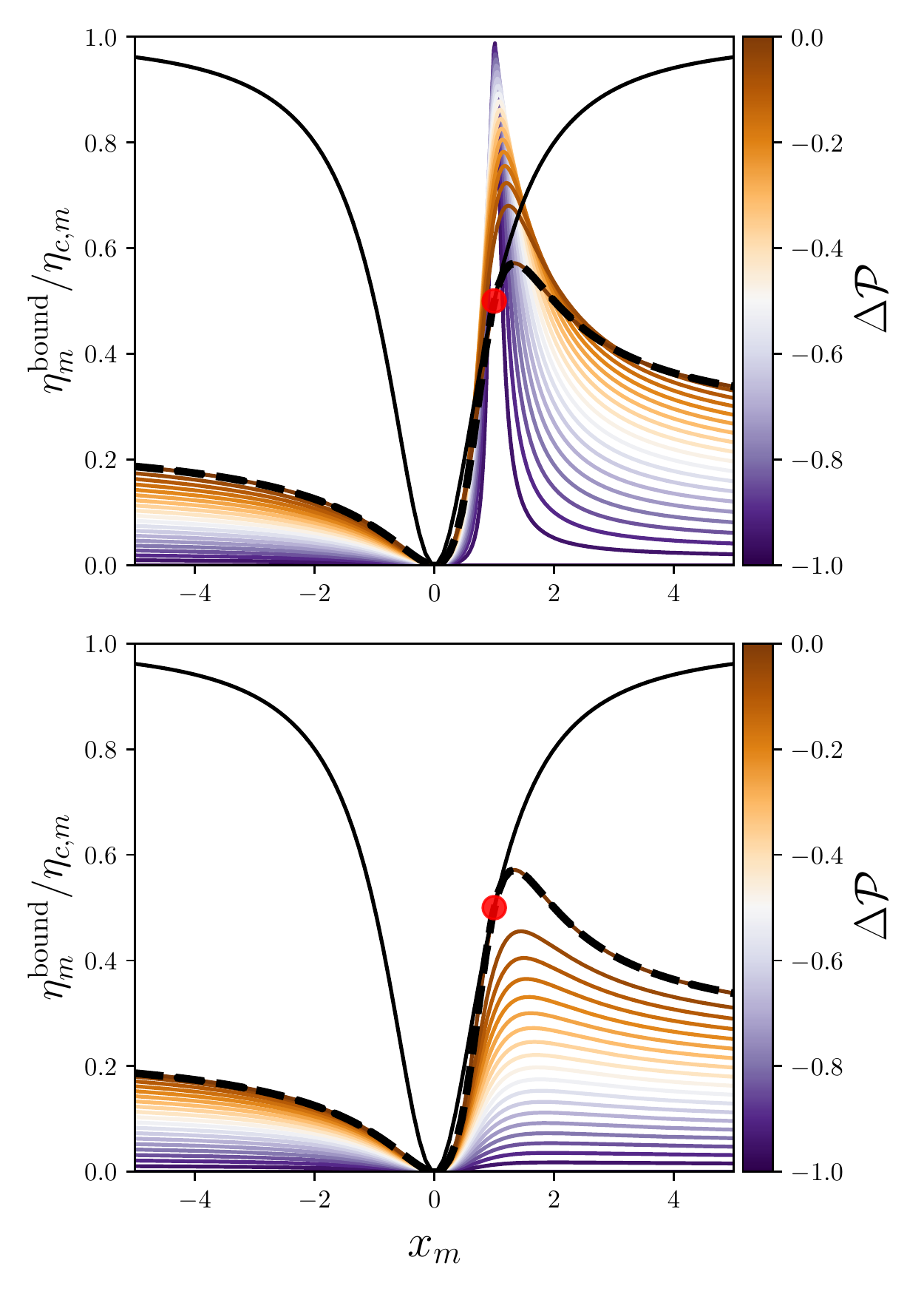}
\caption{Efficiency bound $\eta^{\rm bound}_m/ \eta_{c,m}$ versus the asymmetry parameter $x_m$ and the power gain $\Delta{\cal P}$ for (a): favorable case $\varepsilon_{+}$ and (b): unfavorable case $\varepsilon_{-}$. The intersecting points of the red-circle lines represent the Curzun-Ahlborn limit (CA) efficiency in the linear response regime $\eta_{CA,m}=\eta_{c,m}/2$. For comparison, the bounds obtained by Benenti et al.\cite{a5} only from thermodynamic entropy production $\dot{\cal S}>0$ (black-solid line) and Brandner et al.\cite{a18} on three terminals setup with considering the role of current conservation (black-dashed lines) are included.}
\label{f10}
\end{figure}

Furthermore, the asymmetry parameter $x_m$ is also vital to optimize the performance of a heat engine with broken
time-reversal symmetry at a given output power. In the case that $y_m=H_m$, for different cases expressed in Eqs.~(\ref{e20}) and (\ref{e21}), we can derive the bounds on efficiency $\eta_m$ by means of Eq.~(\ref{e33}) as follows:
\begin{eqnarray}
\eta^{\rm bound}_m=\frac{\eta_{c,m}}{2}\frac{x_m^2(1+\Delta {\cal P})}{2(x_m^2-x_m+1)-(1\pm\sqrt{-\Delta {\cal P}})x_m}.\nonumber\\
\label{e38}
\end{eqnarray}
We noticed that when $J_P^Q=0$, Eq~.(\ref{e38}) becomes equal to the general bound on efficiency obtained by Zhang et al.~\cite{a45} for the B\"uttiker-probe heat engine; thus, the general bounds on efficiency are identical for these two setups. The difference is that for the voltage-probe heat engine, there are three possibilities for the bound efficiency of the heat engine based on Eqs.~(\ref{e20}) and (\ref{e21}). To optimize the general bound on the heat-engine efficiency, we solve the relation $\partial{\eta^{\rm bound}_m}/\partial{x_m}=0$. We found, $x_m=0$ and $x_m=\pm1$ are three possible extreme points. To analyze the dependence of the bound efficiency ($\eta^{\rm bound}_m$) on the $x_m$ as well as $\Delta{\cal P}$, in Fig(\ref{f10}), we display $\eta^{\rm bound}_m$ as a function of $x_m$ for different $\Delta{\cal P}$ and for both favorable and unfavorable case $\varepsilon_{\pm}$, as illustrated in (a) and (b), respectively. From the figure, one can find that the maximum efficiency value is achieved somewhere around the symmetric point ($x_m=1$), and in the limit $x\rightarrow\pm\infty$  functions asymptotically approach the value $\frac{\eta_{c,m}}{4}(1+\Delta{\cal P})$. 
\section{Conclusion}\label{sec4}
Within broken time-reversal symmetry and in the linear response regime, the optimal performance at a given output power of a voltage-probe heat engine is examined. The normalized efficiency at a given output power $\eta_m/\eta_m({\cal P}_{\rm max})$, the asymmetric parameter $x_m$, the general figure of merit $y_m$ and the parameter $d_m$, are first introduced in our work.  $\eta_m/\eta_m({\cal P}_{\rm max})$, $y_m$ and $d_{m}$ are classified into three cases with subscripts $m=L$, $m=P$ and $m=LP$, based on the heat current(s) extracted from the reservoir $L$, reservoir $P$ and both reservoirs $L$ and $P$, respectively. We addressed how a probe reservoir, adjusted to block charge flows and only allow heat flows, can enhance the efficiency at a given output power of the heat engine compared to a B\"uttiker-probe setup, where the probe reservoir blocks both heat and charge flows. In addition, we show that all thermoelectric properties of a B\"uttiker-probe heat engine can be obtained by applying a condition of zero heat flows throughout the probe reservoir and the scattering region. Our results reveal that, unlike the B\"uttiker-probe heat engine, a voltage-probe device due to the parameter $d_m$, can change the upper and lower bounds on $y_m$. Additionally, compared with a B\"uttiker-probe heat engine, where the general figure of merit $y$ and power gain $\Delta{\cal P}$ affect the normalized efficiency at a given output power, for a voltage-probe heat engine the parameter $d_m$ is significantly contributed to change the normalized-efficiency value at the output power, and the optimum value of efficiency increases for $d_m<1$. In other words, the efficiency value at a given output power can be controlled by this key parameter. We also found that, like the Buttiker-probe heat engine, the universal bounds on the efficiency of quantum heat engines with broken time-reversal symmetry. We showed efficiency at a given output power can overcome the Curzon-Ahlborn limit.  One can also extend our theory to a genuine three-terminal heat engine. This setup is more complex than a voltage-probe heat engine, and as a result, it may be beneficial to enhance the normalized efficiency at a given output power compared to voltage-probe or B\"uttiker-probe heat engines.
\appendix
\section{Scattering approach within the linear response regime}\label{apeA}
The coherent flow of heat and particles via a conductor without electron-electron interactions can be defined through the Landauer-B\"uttiker formalism. Taking the assumption that all phase breaking and dissipative processes are restricted to the reservoirs, the heat and particle currents can be written in terms of the scattering characteristics of the scattering region \cite{a46,a47}. For a multi-terminal setup, the particle and heat currents flowing into the scattering region from the $k_{th}$ reservoir can be described using the Landauer-B\"uttiker formalism as follows:
\begin{eqnarray}
J_\alpha^N&=&\frac{1}{h}\int dE\sum_{\alpha\ne\beta}{\cal T}_{\alpha\beta}(E)\Big(f_\alpha(E)-f_\beta(E)\Big),\nonumber\\
J_\alpha^Q&=&\frac{1}{h}\int dE(E-\mu_\alpha)\sum_{\alpha\ne \beta}{\cal T}_{\alpha\beta}(E)\Big(f_\alpha(E)-f_\beta(E)\Big),\nonumber\\
\label{eqA1}
\end{eqnarray}
where $h$ is the Planck’s constant, ${\cal T}_{\alpha\beta}$ is the transmission probability from reservoir $\alpha$ to $\beta$ and $f_\alpha(E)$ is the Fermi function of the respective reservoirs described as 
$f_\alpha(E)=\big[\exp((E-\mu_\alpha)/k_{\rm B}T)+1\big]^{-1}$.
\section{Onsager Coefficients}\label{apeB}
Using Eq.~(\ref{eqA1}), the Onsager coefficients of Eq.~(\ref {e3}) can be obtained from the linear expansion of the currents $J_\alpha^N$ and $J_\alpha^Q$ ($J=\bm{{\cal L}}X$) as follows:
\begin{eqnarray}
{\cal L}_{11}&=&\frac{T}{h}\int_{-\infty}^{\infty} f'(-E)\big({\cal T}_{LP}+{\cal T}_{LR}\big)~dE,
\nonumber\\
{\cal L}_{12}&=&\frac{T}{h}\int_{-\infty}^{\infty} f'(-E) (E-\mu) ({\cal T}_{LP}+{\cal T}_{LR})~dE,
\nonumber\\
{\cal L}_{13}&=&\frac{T}{h}\int_{-\infty}^{\infty} -f'(-E) {\cal T}_{LP}~dE,
\nonumber\\
{\cal L}_{14}&=&\frac{T}{h}\int_{-\infty}^{\infty} f'(-E) (-(E-\mu)) {\cal T}_{LP}~dE,
\nonumber\\
{\cal L}_{21}&=&{\cal L}_{12},
\nonumber\\
{\cal L}_{22}&=&\frac{T}{h}\int_{-\infty}^{\infty} f'(-E)(E-\mu)^2({\cal T}_{LP}+{\cal T}_{LR})~dE,
\nonumber\\
{\cal L}_{23}&=&{\cal L}_{14},
\nonumber\\
{\cal L}_{24}&=&\frac{T}{h}\int_{-\infty}^{\infty} f'(-E)(-(E-\mu)^2){\cal T}_{LP}~dE,
\nonumber\\
{\cal L}_{31}&=&\frac{T}{h}\int_{-\infty}^{\infty} -f'(-E) {\cal T}_{PL}~dE,
\nonumber\\
{\cal L}_{32}&=&\frac{T}{h}\int_{-\infty}^{\infty} f'(-E) (-(E-\mu)) {\cal T}_{PL}~dE,
\nonumber\\
{\cal L}_{33}&=&\frac{T}{h}\int_{-\infty}^{\infty} f'(-E)\big({\cal T}_{PL}+{\cal T}_{PR}\big)~dE
\nonumber\\
{\cal L}_{34}&=&\frac{T}{h}\int_{-\infty}^{\infty} f'(-E)(E-\mu)({\cal T}_{PL}+{\cal T}_{PR})~dE,
\nonumber\\
{\cal L}_{41}&=&{\cal L}_{32},
\nonumber\\
{\cal L}_{42}&=&\frac{T}{h}\int_{-\infty}^{\infty} f'(-E)(-(E-\mu)^2){\cal T}_{PL}~dE,
\nonumber\\
{\cal L}_{43}&=&{\cal L}_{34},
\nonumber\\
{\cal L}_{44}&=&\frac{T}{h}\int_{-\infty}^{\infty} f'(-E)(E-\mu)^2({\cal T}_{PL}+{\cal T}_{PR})~dE,
\nonumber\\
\label{eqB1}
\end{eqnarray}
where $f^{\prime}(E)$ is the derivative of the Fermi-Dirac distribution with respect to the energy. As mentioned before, in a voltage-probe heat engine, the probe reservoir blocks particle flows ($J_{P}^{N}=0$); thus, ${X}_{P}^{V}$ in Eq.~(\ref {e3}) can be derived as follows:
\begin{eqnarray}
{X}_{P}^{V}=\frac{-({\cal L}_{31}{X}_{L}^{V}+{\cal L}_{32}{X}_{L}^T+{\cal L}_{34}{X}_{P}^T)}{{\cal L}_{33}}.
\label{eqB2}
\end{eqnarray}
Setting Eq.~(\ref{eqB2}) into Eq.~(\ref {e3}) , and after some algebras, the Onsager coefficients of Eq.~(\ref {e7}) can be expressed as follows:
\begin{eqnarray}
{\cal L'}_{11}=\frac{{\cal L}_{33}{\cal L}_{11}-{\cal L}_{13}{\cal L}_{31}}{{\cal L}_{33}}&,&
{\cal L'}_{12}=\frac{{\cal L}_{33}{\cal L}_{12}-{\cal L}_{13}{\cal L}_{32}}{{\cal L}_{33}},\nonumber\\
{\cal L'}_{13}=\frac{{\cal L}_{14}{\cal L}_{33}-{\cal L}_{13}{\cal L}_{34}}{{\cal L}_{33}}&,&
{\cal L'}_{21}=\frac{{\cal L}_{21}{\cal L}_{33}-{\cal L}_{23}{\cal L}_{31}}{{\cal L}_{33}},\nonumber\\
{\cal L'}_{22}=\frac{{\cal L}_{33}{\cal L}_{22}-{\cal L}_{23}{\cal L}_{32}}{{\cal L}_{33}}&,&
{\cal L'}_{23}=\frac{{\cal L}_{24}{\cal L}_{33}-{\cal L}_{23}{\cal L}_{34}}{{\cal L}_{33}},\nonumber\\
{\cal L'}_{31}=\frac{{\cal L}_{41}{\cal L}_{33}-{\cal L}_{43}{\cal L}_{31}}{{\cal L}_{33}}&,&
{\cal L'}_{32}=\frac{{\cal L}_{42}{\cal L}_{33}-{\cal L}_{43}{\cal L}_{32}}{{\cal L}_{33}},\nonumber\\
{\cal L'}_{33}=\frac{{\cal L}_{44}{\cal L}_{33}-{\cal L}_{43}{\cal L}_{34}}{{\cal L}_{33}},\nonumber\\
\label{eqB3}
\end{eqnarray}
where ${\cal L}_{ab}$ are given in 
Eq.~(\ref{eqB1}). Analogously, placing $J_{P}^{Q}=0$ into  Eq.~(\ref {e7}), the Onsager coefficients of Eq.~(\ref {eq777}) are expressed as follows:
\begin{eqnarray}
{\cal L}^{{\prime}{\prime}}_{11}=\frac{{\cal L}^{\prime}_{11}{\cal L}^{\prime}_{33}-{\cal L}^{\prime}_{13}{\cal L}^{\prime}_{31}}{{\cal L}^{\prime}_{33}}&,&
{\cal L}^{\prime\prime}_{12}=\frac{{\cal L}^{\prime}_{12}{\cal L}^{\prime}_{33}-{\cal L}^{\prime}_{13}{\cal L}^{\prime}_{32}}{{\cal L}^{\prime}_{33}},\nonumber\\
{\cal L}^{\prime\prime}_{21}=\frac{{\cal L}^{\prime}_{21}{\cal L}^{\prime}_{33}-{\cal L}^{\prime}_{23}{\cal L}^{\prime}_{31}}{{\cal L}^{\prime}_{33}}&,&
{\cal L}^{\prime\prime}_{22}=\frac{{\cal L}^{\prime}_{22}{\cal L}^{\prime}_{33}-{\cal L}^{\prime}_{23}{\cal L}^{\prime}_{32}}{{\cal L}^{\prime}_{33}},\nonumber\\
\label{eqB6}
\end{eqnarray}
where ${\cal L}^{\prime}_{ab}$ are given in Eq.~(\ref{eqB3}).
\section{Carnot Efficiency}\label{apeC}
In any steady-state heat-to-work conversion device, the thermoelectric efficiency is bounded from above by the Carnot engine efficiency $\eta_c$, which can be calculated by placing the condition of zero entropy production rate ($\dot{\cal{S}}$=$0$). For a three-terminal device, it can be derived by analogy with the Carnot efficiency of a two-terminal setup and described in terms of three cases defined in Eqs.~(\ref{e20}) and (\ref{e21}). If the heat current is only absorbed from either reservoir $P$ ($J^Q_P>0$) or $L$ ($J^Q_L>0$), the Carnot efficiency is derived as
\begin{eqnarray}
\eta_{c,L,(P)}={\frac{1}{T}}\Big[\frac{\Delta T_{P}J_{P}^Q+\Delta T_{L}J_{L}^Q}{J_{L,(P)}^{Q}}\Big],
\label{eqA12}
\end{eqnarray}
index $L$ in the denominator is substituted by $P$ if the heat current flows from electrode $L$ into the scattering region. Analogously if heat current is extracted from both reservoirs $P$ and $L$ ($J^{Q}_{L}>0$ and $J^{Q}_{P}>0$), it gives
\begin{eqnarray}
\eta_{c,LP}={\frac{1}{T}}\Big[\frac{\Delta T_{P}J_{P}^Q+\Delta T _{L}J_{L}^Q}{J_{L}^{Q}+J_{P}^{Q}}\Big]
\label{eqA13}
\end{eqnarray}
The Carnot efficiency of the B\"uttiker-probe heat engine  \cite{a45} is obtained by imposing the condition of $J_P^Q=0$. Therefore, for a B\"uttiker-probe device it is expressed as $\eta_{c}=\frac{\Delta T_L}{T}$.
\section{Efficiency at maximum output power with broken time-reversal symmetry}\label{apeD}
Here, we mathematically derive Eqs.~(\ref{e241}). Setting Eqs.~(\ref{e23}) and (\ref{e24}) into the efficiency expressions detailed in Eqs.~(\ref{e20}) and (\ref{e21}), it is possible to obtain efficiency at maximum output power for a voltage-probe engine as follows:
\begin{widetext}
\begin{eqnarray}
\eta_L({\cal P}_{\rm max})&=&\dfrac{1}{2T}\dfrac{\Big[{\cal Z}_L^BT\delta^{-1}+2{\cal Z}_L^AT\Big]{\Delta T}_P+{\cal Z}_L^CT{\Delta T}_L}{\delta^{-1}\Big[{\cal Z}_L^{A^{\prime\prime}}T+2\big(\dfrac{{K}_{LP}}{{K}_{LL}}\big)\Big]+{\cal Z}_{L}^{C^{\prime}}T+2},\\
\eta_P({\cal P}_{\rm max})&=&\dfrac{1}{2T}\dfrac{\Big[{\cal Z}_P^CT\delta+2{\cal Z}_P^AT\Big]{\Delta T}_L+{\cal Z}_P^BT{\Delta T}_P}{\delta\Big[{\cal Z}_P^{A^{\prime}}T+2\big(\dfrac{K_{PL}}{K_{PP}}\big)\Big]+{\cal Z}_P^{B^{\prime}}T+2},\\
\eta_{LP}({\cal P}_{\rm max})&=&\dfrac{1}{2T}\dfrac{\Big[{\cal Z}_{LP}^BT\delta^{-1}+2{\cal Z}_{LP}^AT\Big]{\Delta T}_{P}+{\cal Z}_{LP}^CT{\Delta T}_L}{\delta^{-1}\Big[{\cal Z}_{LP}^{B^{\prime}}T+{\cal Z}_{LP}^{A^{\prime\prime}}T+2\big(\dfrac{{K}_{PP}}{{K}_{LP}}+1\big)\Big]+2\Big(\dfrac{{K}_{PL}+{K}_{LL}}{{K}_{LP}}\Big)+{\cal Z}_{LP}^{A^{\prime}}T+{\cal Z}_{LP}^{C^{\prime}}T},
\label{eqd1}
\end{eqnarray}
\end{widetext}
where ${\cal Z}_m^\theta T$ are the generalized figures of merit for a voltage-probe heat engine, given as follows:
\begin{eqnarray}
{\cal Z}_L^\theta T&=
&\frac{\theta T}{K_{LL}},\nonumber\\
{\cal Z}_P^\theta T&=&\frac{\theta T}{K_{PP}},\nonumber\\
{\cal Z}_{LP}^\theta T&=&\frac{\theta T}{K_{LP}},
\label{eqd2}
\end{eqnarray}
where $\theta=A, A^\prime, A^{\prime\prime}, B, B^\prime, C, C^\prime$ is defined as follows:
\begin{eqnarray}
{A}&=&{S_{LL}(B)S_{LP}(\textbf{B})G_{LL}(\textbf{B})}\nonumber\\
{A^{\prime}}&=&{S_{LL}(\textbf{B})S_{LP}(-\textbf{B})G_{LL}(\textbf{B})}\nonumber\\
{A^{\prime\prime}}&=&{S_{LP}(\textbf{B})S_{LL}(-\textbf{B})G_{LL}(\textbf{B})}\nonumber\\
B&=&{{S_{LP}(\textbf{B})}^2G_{LL}(\textbf{B})}\nonumber\\
{B^{\prime}}&=&{S_{LP}(\textbf{B})S_{LP}(-\textbf{B})G_{LL}(\textbf{B})}\nonumber\\
{C}&=&{{S_{LL}(\textbf{B})}^2G_{LL}(\textbf{B})}\nonumber\\
{C^{\prime}}&=&{S_{LL}(\textbf{B})S_{LL}(-\textbf{B})G_{LL}(\textbf{B})}
\label{eqd3}
\end{eqnarray}
Eq.~(\ref{eqd1}) can also be written in terms of the corresponding Carnot efficiency as follows:
\begin{widetext}
\begin{eqnarray}
\eta_L({\cal P}_{\rm max})&=&\dfrac{\eta_{c,{L}}({\cal P}_{\rm max})}{2}\dfrac{{\cal Z}_L^{C}T{\delta}^2+2\delta{\cal Z}_L^{A}T+{\cal Z}_{L}^{B}T}{\delta T\Big({\cal Z}_{L}^{A^{\prime}}+{\cal Z}_{L}^{A^{\prime\prime}}\Big)+{\cal Z}_{L}^{B^{\prime}}T+{\delta}^2 T{\cal Z}_{L}^{C^{\prime}}+ 2(\dfrac{\delta K_{PL}+K_{LP}}{K_{LL}})+2\Big(\dfrac{K_{PP}}{K_{LL}}\Big)+2{\delta}^2}\nonumber\\
&=&\dfrac{\eta_{c,L}({\cal P}_{\rm max})}{2}\frac{r_L}{y_L+2d_L}\\
\eta_P({\cal P}_{\rm max})&=&\dfrac{\eta_{c,{P}}({\cal P}_{\rm max})}{2}\dfrac{{\cal Z}_{P}^{C}T\delta^2+2\delta{\cal Z}_{P}^{A}T+{\cal Z}_P^{B}T}{\delta T\Big({\cal Z}_{P}^{A^{\prime}}+{\cal Z}_P^{A^{\prime\prime}}\Big)+{\cal Z}_{P}^{B^{\prime}}T+{\delta}^2 T{\cal Z}_P^{C^{\prime}}+ 2\delta\Big(\dfrac{K_{PL}+K_{LP}}{K_{PP}}\Big)+2{\delta}^2\Big(\dfrac{K_{LL}}{K_{PP}}\Big)+2}\nonumber\\
&=&\dfrac{\eta_{c,{P}}({\cal P}_{\rm max})}{2}\frac{r_P}{y_P+2d_P}\\
\eta_{LP}({\cal P}_{\rm max})&=&\dfrac{\eta_{c,{LP}}({\cal P}_{\rm max})}{2}\dfrac{{\cal Z}_{LP}^{C}T\delta^2+2\delta{\cal Z}_{Lp}^{A}T+{\cal Z}_{LP}^{B}T}{\delta T\Big({\cal Z}_{LP}^{A^{\prime}}+{\cal Z}_{LP}^{A^{\prime\prime}}\Big)+{\cal Z}_{LP}^{B^{\prime}}T+{\delta}^2 T{\cal Z}_{Lp}^{C^{\prime}}+2\Big(\dfrac{\delta K_{PL}+K_{PP}}{K_{LP}}\Big)+2{\delta}^2\Big(\dfrac{K_{LL}}{K_{LP}}\Big)+2\delta}\nonumber\\
&=&\dfrac{\eta_{c,{LP}}({\cal P}_{\rm max})}{2}\frac{r_{LP}}{y_{LP}+2d_{LP}}
\label{eqA5}
\end{eqnarray}
\end{widetext}
After some algebras, maximum output power efficiency for a voltage-probe heat engine can be obtained as follows: 
\begin{eqnarray}
\eta_m({\cal P}_{\rm max})=\frac{\eta_{c,m}({\cal P}_{\rm max})}{2}\frac{x_my_m}{y_m+2d_m}&;& m=L, P, LP\nonumber\\
\label{eqA6}
\end{eqnarray}
where $x_m=\dfrac{r_m}{y_m}$.
\begin{figure*}[t]
\includegraphics[width=1\columnwidth]{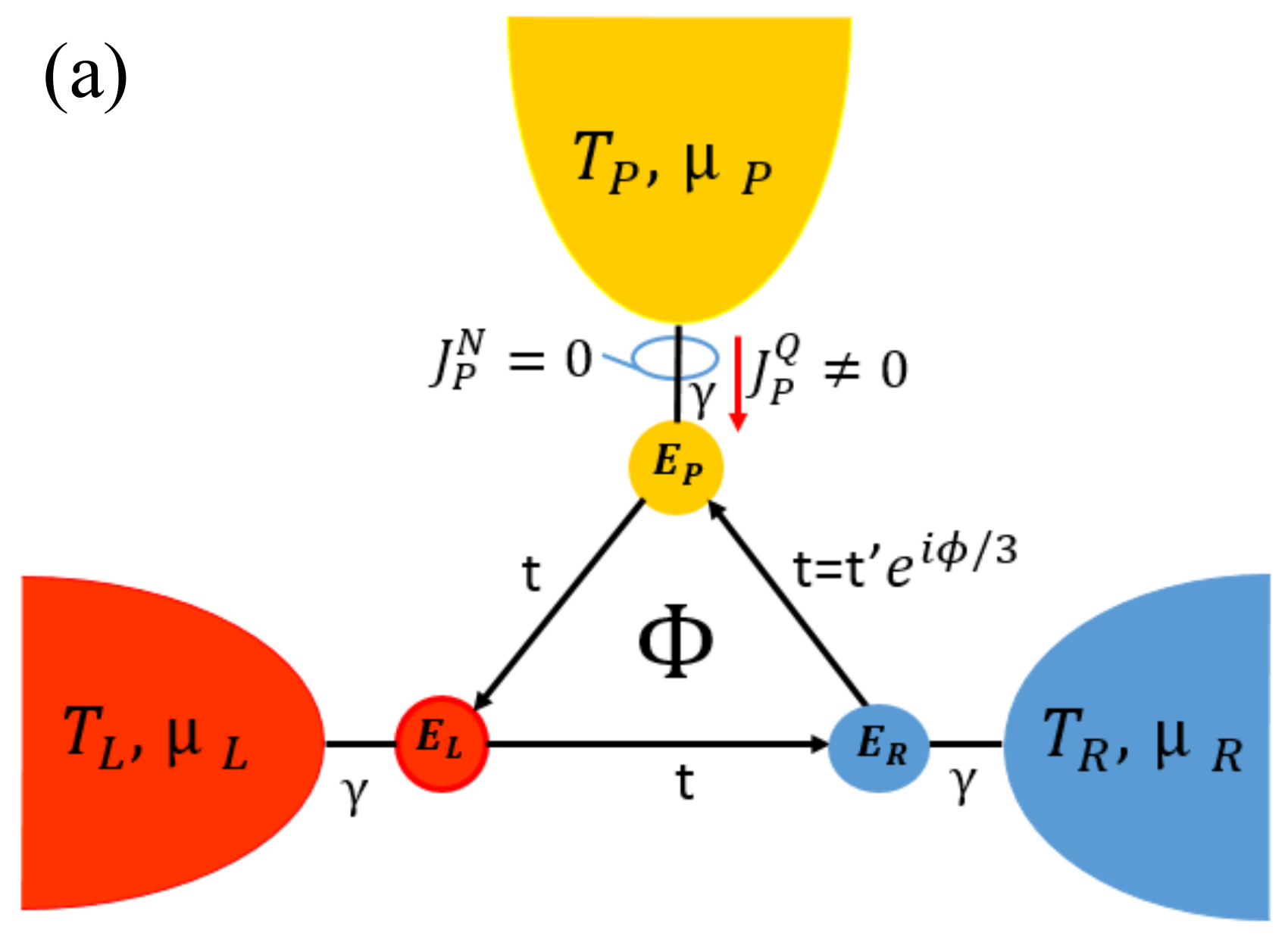}
\includegraphics[width=2\columnwidth]{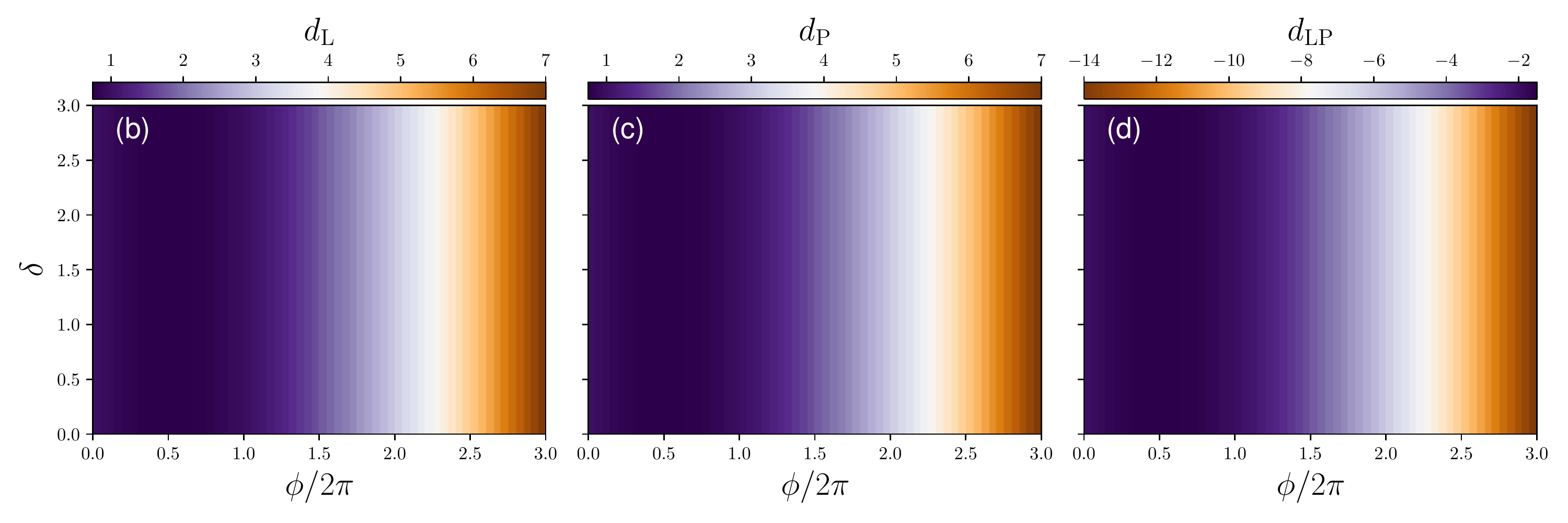}
\caption{(a) shows a voltage-probe system consisting of a triple quantum dot with a single energy level, subjected to the external magnetic field $(\textbf{B})$ and connected to three electronic reservoirs, labeled $L$, $P$ and $R$, in different temperatures $(T_{L}>T_{P}>T_{R})$ and chemical potentials, $\mu_{L}$, $\mu_{P}$ and $\mu_{R}$. (b), (c) and (d) exhibit the contour plots of $d_L$,  $d_{P}$ and $d_{LP}$, respectively, as a function of $\phi$ and $\delta$ for the system illustrated in (a). The atomic site energy of the quantum dots connected to the reservoirs $L, P$ and  $R $ are $E_{L}-{\mu}= 1.0~k_BT$, $E_P-{\mu}=1.0~k_BT$ and $E_R-{\mu}=1.0~k_BT$, respectively, where $\mu=\mu_R=0$, $T=T_R$ and $k_BT=1$. Coupling parameter: $\gamma_{L}=\gamma_{P}=\gamma_{R}=\gamma=0.5~k_BT$. Hopping energy parameter between two atomic sites: $t=t'e^{i\phi/3}$, where $\phi=2\pi
\Phi/\Phi_0$ ($\Phi_0$ is quantum flux ) and $t'=1.0~k_BT$.}
\label{fig_appx}
\end{figure*}
\section{Toy model}\label{apeE}
We consider a toy quantum heat engine in the presence of an external magnetic field to investigate numerically the sign and the range of the parameter $d_m$, see Eq.~(\ref{e27}). Fig.~\ref{fig_appx}-(a) depicts a voltage-probe engine consisting of a triple quantum dot with a single energy level, subjected to an external magnetic field $\textbf{B}$ which are connected to three electronic reservoirs, labeled $L$, $P$ and $R$, in different temperatures $(T_L>T_P>T_R)$ and chemical potentials,  $\mu_L$, $\mu_P$ and $\mu_R$. For simplicity, the coupling strength to reservoirs $L$, $P$, and $R$ are taken equal to $\gamma$, and $E_\alpha$ ($\alpha=L, P, R$) denotes the atomic site energy of respective quantum dots. To describe the electronic structure as well as the transport properties of the setup under consideration, we utilize the non-equilibrium Green function technique \cite{a40}. Once the Transmission function is obtained, through the Onsager matrix, expressed in Eq.~(\ref{e7})), one can find all driven currents in the setup.
\par
Figs.~\ref{fig_appx}-(b), (c), (d) demonstrate the plots of $d_L$, $d_P$ and $d_{LP}$ as a function of $\phi$ and $\delta$. It can be seen that $d_L$ and $d_P$ are positive as a function of $\delta$ and $\phi$, while $d_{LP}$ is negative. It is evident that by tuning the ratio $\delta$= ${X_L^T}/{X_L^P}$ and $\phi$, we can tune the value of the parameter $d_m$ in our setup.

\bibliography{references/ref}

\end{document}